\documentclass[draft]{article}
\usepackage{amsmath,amsfonts,amsthm,amssymb,amscd}

\newcommand{\be}{\begin{equation}}
\newcommand{\ee}{\end{equation}}
\newcommand{\bs}{\begin{split}}
\newcommand{\es}{\end{split}}
\newcommand{\ba}{\begin{align}}
\newcommand{\ea}{\end{align}}

\theoremstyle{plain} \newtheorem{theorem}{Theorem}[section]
\newtheorem{lemma}[theorem]{Lemma}
\newtheorem{proposition}[theorem]{Proposition}
\newtheorem{corollary}[theorem]{Corollary} %\theoremstyle{definition}
\newtheorem{remark}[theorem]{Remark}
\newtheorem{hypo}[theorem]{Hyp.}

\newcommand{\R}{{\mathbb R}}

\newcommand{\F}{{\mathcal F}}
\newcommand{\C}{\mathbb{C}}
 
\newcommand{\pp}{P_{3}}
\newcommand{\esp}{E_{\sigma}(P_{3})}
\newcommand{\psp}{\Phi_{\sigma}(P_{3})}

\newcommand{\hsp}{H_{\sigma}(P_{3})}
\newcommand{\hop}{H_{0}(P_{3})}
\newcommand{\hisp}{H_{I,\sigma}(P_{3})}
\newcommand{\ths}{\tilde H_{\sigma}}
\newcommand{\tho}{\tilde H_{0}}
\newcommand{\this}{\tilde H_{I,\sigma}}
\newcommand{\emsp}{E_{\mathrm{mod},\sigma}(P_{3})}
\newcommand{\hmsp}{H_{\mathrm{mod},\sigma}(P_{3})}
\newcommand{\htmsp}{\check{H}_{\mathrm{mod},\sigma}(P_{3})}
\newcommand{\omod}{\omega_{\mathrm{mod}}(k)}
\newcommand{\omo}{\omega_{\mathrm{mod}}}
\newcommand{\om}{\Omega_{ph}}
\newcommand{\rs}{\rho_{\sigma}}

\newcommand{\va}[1]{|#1|}
  \newcommand{\vak}{|k|}
  \newcommand{\vaka}{|k_{3}|}

  \newcommand{\vag}{|g|}
  
\newcommand{\norm}[1]{\|#1\|}

\numberwithin{equation}{section}

\begin{document}

\author{
%------------------
Laurent AMOUR,%
\footnote{Laboratoire de Math\'ematiques, UMR-CNRS 6056,
Universit\'e de Reims,
 Moulin de la Housse - BP 1039, 
 51687 REIMS Cedex 2, France. }
%-------------------
\quad  Beno\^\i t GR\'EBERT%
\footnote{Laboratoire de Math\'ematiques Jean LERAY, UMR-CNRS 6629,
Universit\'e de
Nantes, 2, rue de la Houssini\`ere, 44072 NANTES Cedex 03, France.}
%------------------
\\ and \\ Jean-Claude GUILLOT%
%\footnote{D\'epartement de Math\'ematiques, UMR-CNRS 7539, Institut 
%Galil\'ee, Universit\'e Paris-Nord, 93430 Villetaneuse, France.}
\footnote{ Centre de Math\'ematiques Appliqu\'ees, UMR-CNRS 7641, Ecole 
Polytechnique, 91128 Palaiseau Cedex, France.}
}

\title{The dressed nonrelativistic  electron\\ in a magnetic 
field}

\date{ }
\maketitle

\begin{abstract}
    We consider a nonrelativistic electron interacting 
    with a classical magnetic field pointing along the $x_{3}$-axis and 
    with a quantized electromagnetic field. When the interaction 
    between the electron and photons is turned off, the electronic 
    system is assumed to have a ground state of finite multiplicity.
    Because of the translation 
    invariance along the $x_{3}$-axis, we consider the reduced Hamiltonian 
    associated with the total momentum along the $x_{3}$-axis and, after 
    introducing an ultraviolet cutoff and an infrared regularization, we 
    prove that the reduced Hamiltonian has a ground state if the coupling 
    constant and the total momentum along the $x_{3}$-axis are 
    sufficiently small. We determine the absolutely continuous 
    spectrum of the reduced Hamiltonian and, when the ground state is 
    simple, we prove that the renormalized mass of the dressed 
    electron is greater than or equal to its bare one. We then deduce that the 
    anomalous magnetic moment of the dressed electron is non negative.
    
    {\textbf{ AMS}} classification numbers: \textbf{81V10, 81Q10, 81Q15} 
\end{abstract}

\newpage
\tableofcontents
\newpage
\setcounter{section}{0}

\section{Introduction}

We consider a nonrelativistic electron in $\R^3$ of charge $e$ and mass $m$ 
interacting with a magnetic field pointing along the $x_{3}$-axis and 
with photons.
The magnetic field takes the form 
$(0,0,b(x_{1},x_{2}))$ with
$b(x_{1},x_{2})= \frac{\partial a_{2}}{\partial x_{1}}(x_{1},x_{2})-
\frac{\partial a_{1}}{\partial x_{2}}(x_{1},x_{2})$ where 
$a(x_{1},x_{2})$ is a vector potential. The associated Pauli 
Hamiltonian in Coulomb gauge is formally given by
\begin{align}
      \begin{split}\label{Hfor}
    H=&\frac{1}{2m }(p-ea(x')-eA(x))^{2} 
    -\frac{e}{2m}b(x'){\sigma}_{3}\otimes 1 + V(x') \otimes 1 \\+& 1 \otimes 
    H_{ph} -\frac{e}{2m}\sigma \cdot B(x) \ .
\end{split}
   \end{align}
Here the units are such that $\hbar =c=1$, $p=-i\nabla_{x}$,
$x=(x_{1},x_{2},x_{3}
)$ together with $x'=(x_{1},x_{2})$, $\sigma =(\sigma_{1},\sigma_{2},\sigma_{3})$
is the 3-component vector of the Pauli 
matrices and $V(x')$ is an electric potential depending only on the 
transverse variables.
The quantized  electromagnetic field in the Coulomb gauge is formally  given by 
\be
A(x)= \frac{1}{2\pi}\sum_{\mu =1,2} \int d^3k \left( 
\frac{1}{\vak^{1/2}}\epsilon_{\mu}(k)e^{-ik\cdot 
x}a^\star_{\mu}(k)\right. 
+ \left. \frac{1}{\vak^{1/2}}\epsilon_{\mu}(k)e^{ik\cdot 
x}a_{\mu} (k)\right)\ ,
\ee
\begin{align}
      \begin{split}
B(x)= \frac{i}{2\pi}\sum_{\mu =1,2} \int d^3k& \left\{ 
-\vak^{1/2}\left(\frac{k}{\vak}\wedge\epsilon_{\mu}(k)\right)\right.
e^{-ik\cdot 
x}a^\star_{\mu}(k) \\
&+  \left. \vak^{1/2}\left(\frac{k}{\vak}\wedge\epsilon_{\mu}(k)\right)
e^{ik\cdot 
x}a_{\mu}(k)\right\}
\end{split}
  \end{align}
 where 
$\epsilon_{\mu}(k)$ are real polarization vectors 
satisfying $\epsilon_{\mu}(k)\cdot \epsilon_{\mu'}(k)=\delta_{\mu \mu'}$,
$k\cdot
\epsilon_{\mu}(k)=0$; $a_{\mu}(k)$ and $a^\star_{\mu}(k)$ are the 
usual annihilation and creation operators acting in the Fock space
$$
\mathcal F := \oplus_{n=0}^{\infty} L^2(\R^{3},\C^2)^{\otimes^n_{s}}
$$
where $L^2(\R^{3},\C^2)^{\otimes^0_{s}}=\C$ and $L^2(\R^{3},\C^2)^{\otimes^n_{s}}$
is the symmetric $n$-tensor power of $L^2(\R^{3},\C^2)$ appropriate 
for Bose-Einstein statistics.
The annihilation and creation operators  obey the canonical 
commutation relations ($a^\sharp=a^\star$ or $a$) 
\be \label{ccr}
[a^\sharp_{\mu}(k),a^\sharp_{\mu'}(k') ]=0 \quad \mbox{et} \quad
[a_{\mu}(k),a^\star_{\mu'}(k') ]=\delta_{\mu \mu'}\delta(k-k')\ .
\ee
Finally the Hamiltonian for the photons is given by
\be \label{Hphoton}
H_{ph}= \sum_{\mu=1,2}\int \vak 
a_{\mu}^\star(k) a_{\mu}(k) d^3k\ .
\ee
The Hilbert space associated with $H$ is then
$$\mathcal H = L^2(\R^3, \C^2)\otimes \mathcal F \simeq
L^2(\R^3, \C^2\otimes \mathcal F)\ .$$
As it stands, the Hamiltonian $H$ cannot be defined as a self-adjoint 
operator in $\mathcal H$ and we need to introduce cutoff functions, 
both in $A(x)$ and in $B(x)$, which will satisfy appropriate 
hypothesis in order to get a self adjoint operator in $\mathcal H$.

\medskip

This operator, still denoted by $H$, commutes with the third 
component, denoted by $P_{3}$, of the total momentum of the system 
(cf. \cite{AHS}). We have $P_{3}= p_{3}\otimes 1 + 1 \otimes {\rm d}\Gamma (k_{3})$
where ${\rm d}\Gamma (k_{3})$ is the second quantized operator 
associated to the multiplication operator by the third component of 
$k$ in $L^2(\R^3,\C^2)$. The spectrum of $P_{3}$ is the real line. In 
turns out that $H$ admits a decomposition over the spectrum of $P_{3}$ 
as a direct integral
\be 
H \simeq \int_{\R}^{\oplus}H(P_{3}) dP_{3}
\ee
on
$$
\mathcal H \simeq \int_{\R}^{\oplus}L^2(\R^2, \C^2\otimes \mathcal F 
)d P_{3}
\simeq \int_{\R}^{\oplus}L^2(\R^2, \C^2)\otimes \mathcal F dP_{3}\ .
$$
The reduced operator $H(P_{3})$ will be explicitely computed and the 
aim of this article is to initiate the spectral analysis of 
$H(P_{3})$ when $\va{P_{3}}$ is small. The results of this work 
were announced in \cite{AGG05a}.

In the free case, i.e. when $b=V=0$, a similar problem has been studied 
in \cite{Chen} by T. Chen who considers a freely propagating nonrelativistic 
spinless charged particle interacting with the quantized 
electromagnetic field.  Under an infrared 
regularization hypothesis, T. Chen proves that the reduced Hamiltonian 
associated to the total momentum $P$ has a unique ground state when 
$P$ is sufficiently small by applying the renormalization group method 
introduced in \cite{BFS98a} (see also \cite{HI04}). In the case of the one-particle sector of Nelson's 
model, a similar result has been obtained first by J. Fr\"ohlich (see 
\cite{F74}, \cite{F73}) and more 
recently by A. Pizzo (see \cite{piz03}, \cite{piz04}) and J.S. 
M{\o}ller \cite{Mo}. For a review of the mathematical problems of nonrelativistic quantum
electrodynamics see \cite{Hiro}.

The hypothesis about $b$ and $V$ (see section \ref{pauli} for the precise 
assumption) are such that the electronic part of the Hamiltonian 
\ref{Hfor} has a finitely degenerated ground state and we face the 
problem of perturbing an eigenvalue of finite multiplicity at the 
bottom of an essential spectrum. 
In this paper we give a simple proof for the 
existence of a ground state for $H(P_{3})$ with an ultraviolet cutoff 
and an infrared regularization. The proof borrows ideas both from 
\cite{F74} (see also \cite{F73,piz03,piz04,FGSray}) 
where the Hamiltonian is invariant by translation and 
\cite{BFS99} (see also \cite{BCV,H01,H99,H00,H04,GLL,LL03,FGScom}
where the electronic part is confined).

When the ground state is simple we are able to prove that the 
renormalized mass of the electron is greater than or equal to its bare one and we 
deduce that the anomalous magnetic moment of the electron is non 
negative.

The question of removing the infrared regularization in such QED 
models for one electron is still open.
Following \cite{Chen}, we can conjecture that, for $P_{3} \neq 0$, $H(P_{3})$ 
without infrared regularization has no ground state in $\mathcal H$ 
(actually the ground state should leave the Fock space when the 
infrared regularization is turned off). 

The same proof also works for any free atom or positive ion  
interacting with the quantized electromagnetic field. Furthermore, in 
the neutral case, i.e. in the case of atoms, we can remove the 
infrared regularization by using a Power-Zienau-Woolley 
transformation (see 
\cite{AGG05b}). 

\vskip20pt\noindent {\it Acknowledgements} LA and BG acknowledge the 
hospitality of the Centre de Math\'ematiques Appliqu\'ees at the \'Ecole 
Polytechnique where a large part of this work has been done.

\section{Definition of the model and self-adjointness}

The Hamiltonian $H$ can be written as
\be \label{H}
H=H_{0}+H_{I}
\ee
where
\begin{eqnarray}\label{H0}
    H_{0}=\left\{\frac{1}{2m }p_{3}^2 +\frac{1}{2m}\sum_{j=1,2}(p_{j}-ea_{j}(x'))^2 
    -\frac{e}{2m}b(x')\sigma_{3}+ V(x')\right\}\otimes 1 + 1 \otimes 
    H_{ph} 
\end{eqnarray}
and $H_{I}$ describes the interaction between the electron in the 
magnetic field $b(x')$ with the photons. A basic tool is now to 
describe the spectral properties of the Pauli operator in 
$L^2(\R^2,\C^2)$ that we are dealing with.

\subsection{The Pauli operator with magnetic 
fields}\label{pauli}

Let $h(b,V)$ be the following operator in $L^2(\R^2,\C^2)$
\be \label{hbV}
h(b,V)=\frac{1}{2m}\sum_{j=1,2}(p_{j}-ea_{j}(x_{1},x_{2}))^2 
    -\frac{e}{2m}b(x_{1},x_{2})\sigma_{3}+ V(x_{1},x_{2}) \ .
\ee
As in \cite{IT98} the $a_{j}$'s are real functions in $C^1(\R^2)$ 
such that
$$b(x_{1},x_{2})=\frac{\partial}{\partial 
x_{1}}a_{2}(x_{1},x_{2})-\frac{\partial}{\partial 
x_{2}}a_{1}(x_{1},x_{2})\ .$$
We suppose that $b$ and $V$ satisfy the 
following hypothesis:

\begin{hypo}\label{H1}
$b$ and $V$ are such that $h(b,V)$ is essentially self adjoint  
on $C_{0}^\infty 
(\R^2,\C^2)$ and  the bottom of the spectrum of $h(b,V)$ is a 
strictly negative isolated eigenvalue of finite multiplicity.
\end{hypo}

There exist many examples of $b$ and $V$ satisfying the hypothesis 
\ref{H1}. Let us give one example.

Suppose that   $b\in C^{1}(\R^2)$ and  $V\in L^\infty (\R^2)$ 
sastisfy
     \be \label{h1}
     1/C \leq b(x_{1},x_{2})\leq C \mbox{ and } \va{\nabla 
     b(x_{1},x_{2})}\leq C
     \ee
     for some $C>1$ and
     \be \label{h2}
     V(x_{1},x_{2})\to 0 \mbox{ as } \va{(x_{1},x_{2})}\to +\infty \ .
    \ee 
    
We then have

\begin{proposition}
Suppose that $b$ and $V$ satisfy \eqref{h1} and \eqref{h2}. Then the operator 
$h(b,V)$ with domain 
$$
D(b,V)=\left\{u\in L^2(\R^2, \C^2)\mid h(b,V)u\in L^2(\R^2, 
\C^2)\right\}
$$
is self-adjoint in $L^2(\R^2, \C^2)$.

Furthermore $h(b,V)$ is essentially self-adjoint on $C_{0}^\infty 
(\R^2,\C^2)$.
\end{proposition}
The proof can be found in \cite{AHS78b} (see also \cite{DR04} and 
\cite{GL02}).

According to \cite{Shi91} (see also \cite{CFKS,Sob96,Rai99})
$\frac{1}{2m}\sum_{j=1,2}(p_{j}-ea_{j}(x_{1},x_{2}))^2 
-\frac{e}{2m}b(x_{1},x_{2})\sigma_{3}$ has zero as an eigenvalue 
of infinite multiplicity when $b$ satisfies \eqref{h1}. By 
adding $V(x_{1},x_{2})$ satisfying \eqref{h2}, the operator $h(b,V)$  may have 
eigenvalues of finite multiplicity accumulating at zero. In fact 
according to \cite{IT98,Rai99} there 
exist  $b$ and $V$ satisfying \eqref{h1} and \eqref{h2} such that 
hypothesis \ref{H1} is verified.

Notice that hypothesis \eqref{H1} allows us to choose an uniform 
magnetic field but, in this case, $V$ cannot be identically zero.

\subsection{The model}
We now introduce the Hamiltonian in the Fock space associated 
to \eqref{Hfor}. As usual we will consider the charge $e$ in front of 
the quantized electromagnetic field $A(x)$ as a parameter that from now 
on  we 
denote  by $g$. 

We introduce  $\rho (k)$ a cutoff function 
associated with an
ultraviolet cutoff 
%($\rho (k)=0$ for $\vak \geq \Lambda$ for some 
%arbitrary $\Lambda$) 
and an infrared regularization. The precise  assumption
verified by $\rho$ will be given in each 
theorem but $\rho$ will always satisfy \eqref{rho} below.

The associated quantized electromagnetic field is then given by 
($j=1,2,3$)
\begin{eqnarray} \label{Aj}
A_{j}(x, \rho)= \frac{1}{2\pi}\sum_{\mu =1,2} \int d^3k \left( 
\frac{\rho(k)}{\vak^{1/2}}\epsilon_{\mu}(k)_{j}e^{-ik\cdot 
x}a^\star_{\mu }(k)\right. \\ \nonumber 
+ \left. \frac{\bar{\rho}(k)}{\vak^{1/2}}\epsilon_{\mu}(k)_{j}e^{ik\cdot 
x}a_{\mu} (k)\right)\ ,
\end{eqnarray}
\begin{eqnarray} \label{Bj}
B_{j}(x, \rho)= \frac{i}{2\pi}\sum_{\mu =1,2} \int d^3k \left( 
-\vak^{1/2}\rho(k)\left(\frac{k}{\vak}\wedge\epsilon_{\mu}(k)\right)_{j}
e^{-ik\cdot 
x}a^\star_{\mu} (k)\right.\\ \nonumber
+ \left. \vak^{1/2}\bar{\rho}(k)\left(\frac{k}{\vak}\wedge
\epsilon_{\mu}(k)\right)_{j}
e^{ik\cdot 
x}a_{\mu} (k)\right)\ .
\end{eqnarray}

The interaction Hamiltonian (cf. \eqref{H}) reads
\begin{align}\nonumber
    H_{I}=& -\frac{g}{m}\sum_{j=1,2}
    \left\{A_{j}(x,\rho)(p_{j}-ea_{j}(x'))+
    (p_{j}-ea_{j}(x'))A_{j}(x,\rho)\right\}\\ \nonumber
    &-\frac{g}{m }\left\{A_{3}(x,\rho)p_{3} 
    +p_{3}A_{3}(x,\rho)\right\}\\ \nonumber
     &-\frac{g}{2m}\sigma \cdot B(x,\rho)+\frac{g^2}{2m}
     A(x,\rho)\cdot A(x,\rho)\ .
\end{align}
Noticing that $k\cdot \epsilon_{\mu}(k) =0$, $H_{I}$ can be 
rewritten as
\begin{align}\begin{split}\label{HI}
    H_{I}=&-\frac{g}{m }A_{3}(x,\rho)p_{3} -\frac{g}{m}\sum_{j=1,2}
    A_{j}(x,\rho)(p_{j}-ea_{j}(x'))\\
    &-\frac{g}{2m}\sigma \cdot B(x,\rho)+\frac{g^2}{2m}
     :A(x,\rho)\cdot A(x,\rho):
\end{split}\end{align}
where we also substitute the Wick normal ordering $:A(x,\rho)\cdot 
A(x,\rho):$ for $A(x,\rho)\cdot A(x,\rho)$. This  substitution 
changes the Hamiltonian by a constant as it follows from the 
canonical commutation relations.

Let $\mathcal F_{0,fin}$ be the set of $(\psi_{n})_{n\geq 0} \in 
\mathcal F$ such that $\psi_{n}$ is in the Schwartz space for every 
$n$ and $\psi_{n}=0$ for all but finitely many $n$.
Suppose that 
\be \label{rho}
\int_{\vak \leq 1} \frac{\va{\rho(k)}^{2}}{\vak^2}d^3k <\infty \quad
    \mbox{and}\quad 
    \int_{\vak \geq 1} {\vak}{\va{\rho(k)}^2}d^3k <\infty \ .
\ee
Then our model is described by the operator $H=H_{0}+H_{I}$
 given by \eqref{H0} and \eqref{HI}, and this operator is well defined on 
 $C^\infty_{0}(\R^3,\C^2)\otimes \mathcal F_{0,fin}$.

\subsection{Self-adjointness}\label{sa}
In $L^2(\R^3,\C^2)\otimes \mathcal F$, the operator $H_{0}$
 given by \eqref{H0} is essentially self adjoint on 
 $C^\infty_{0}(\R^3,\C^2)\otimes \mathcal F_{0,fin}$ (see \cite{RS2}). Its 
 self-adjoint extension is still denoted by $H_{0}$. The interaction 
 operator $H_{I}$ (see \eqref{HI}) is a symmetric operator on
 $C^\infty_{0}(\R^3,\C^2)\otimes \mathcal F_{0,fin}$. We are going to prove 
 that $H_{I}$ is relatively bounded with respect to $H_{0}$ to apply 
 the Kato-Rellich theorem.
 \begin{theorem}\label{thm2}
Assume \eqref{rho}
and
\be \nonumber
\frac{6\va{g}}{\pi \sqrt{2m}}\left(\int \frac{\va{\rho(k)}^{2}}{\vak^2}d^3k \right)^{1/2} +
\frac{g^2}{\pi^2m}\int \frac{\va{\rho(k)}^{2}}{\vak^2}d^3k    <\frac{1}{2}\ .
\ee    
Then $H$ is a self-adjoint operator in $\mathcal H$ with domain 
$D(H)=D(H_{0})$ and $H$ is essentially self-adjoint on
$C^\infty_{0}(\R^3,\C^2)\otimes \mathcal F_{0,fin}$.
\end{theorem} 
\proof To begin with we recall the following well known estimates 
(cf. \cite{BFS98b})
\be\label{aa}
\norm{a_{\mu}(g(.,x))\psi}\leq \left(\int \frac{\va{g(x,k)}^2}{ 
\vak}d^3k\right)^{1/2}\norm{(I\otimes H_{ph}^{1/2})\psi}
\ee
and
\begin{align}\begin{split}\label{aaa}
\norm{a_{\mu}^*(g(.,x))\psi}&\leq \left(\int \frac{\va{g(x,k)}^2}{ 
\vak}d^3k\right)^{1/2}\norm{(I\otimes H_{ph}^{1/2})\psi}\\ 
&+\left(\int {\va{g(x,k)}^2}d^3k\right)^{1/2}\norm{\psi}\ .
\end{split}\end{align}
 We get for $\psi \in 
C^\infty_{0}(\R^3,\C^2)\otimes \mathcal F_{0,fin}$ 
\begin{align}\begin{split}\nonumber
\frac{\vag}{m}\norm{A_{3}(x,\rho)p_{3}\psi}&\leq
\frac{4\vag}{\pi\sqrt{2m}}\left( \int 
\frac{\va{\rho(k)}^2}{\vak^2}d^3k\right)^{1/2}\\ &\norm{ (I\otimes 
H_{ph}^{1/2})(\frac{p_{3}}{\sqrt{2m}}\otimes I)\psi}\\
+\frac{2\vag}{\pi\sqrt{2m}}\left( \int 
\frac{\va{\rho(k)}^2}{\vak}d^3k\right)^{1/2}& \norm{ 
( \frac{p_{3}}{\sqrt{2m}}\otimes I)\psi}\ .
\end{split}\end{align}
Denoting $e(b,V):= \inf \sigma (h(b,V))$, notice that
\begin{align} \nonumber
\norm{(I\otimes 
H_{ph}^{1/2})( \frac{p_{3}}{\sqrt{2m}}\otimes I)\psi}&\leq \frac{1}{2}
\left(\norm{(I\otimes H_{ph})\psi}+ 
\norm{(\frac{p_{3}}{\sqrt{2m}}\otimes I)\psi} \right)\\ \nonumber
&\leq \norm{(H_{0}-e(b,V))\psi}
\end{align}
and, for every $\epsilon >0$,
$$
\norm{( \frac{p_{3}}{\sqrt{2m}}\otimes I)\psi}\leq 
\sqrt\frac{\epsilon}{2}\norm{(H_{0}-e(b,V))\psi}+ 
\frac{1}{\sqrt{2\epsilon}}\norm{\psi}
$$
to obtain that 
\begin{align}\begin{split}\label{sa1}
\frac{\vag}{m}\norm{A_{3}(x,\rho)p_{3}\psi}&\leq
\frac{4\vag}{\pi\sqrt{2m}}\left( \int 
\frac{\va{\rho(k)}^2}{\vak^2}d^3k\right)^{1/2}\norm{(H_{0}-e(b,V)) 
\psi} \\ + \frac{2\vag}{\pi\sqrt{2m}}&\left( \int  
\frac{\va{\rho(k)}^2}{\vak}d^3k\right)^{1/2}
\left(\sqrt\frac{\epsilon}{2}\norm{(H_{0}-e(b,V))\psi}+ 
\frac{1}{\sqrt{2\epsilon}}\norm{\psi}\right)\ .
\end{split}
\end{align}
Therefore $\frac{g}{m}A_{3}(x,\rho)p_{3}$ is relatively bounded 
with respect to $H_{0}$ with relative bound $
\frac{4\vag}{\pi\sqrt{2m}}\left( \int 
\frac{\va{\rho(k)}^2}{\vak^2}d^3k\right)^{1/2}$.
Similarly we verify that for $j=1,2$,  $
\frac{g}{m}A_{j}(x,\rho)(p_{j}-ea_{j}(x'))$ is also relatively bounded 
with respect to $H_{0}$ with the same relative bound, 
 and  that $\frac{\vag}{2m} \sigma \cdot B(x,\rho)$ is relatively bounded 
with respect to $H_{0}$ with a zero relative bound.

It remains to estimate the quadratic terms associated with
$\frac{g^2}{2m}
     :A(x,\rho)\cdot A(x,\rho):$. Let us recall the following 
     estimates (cf. \cite{A90}):

\begin{align} \begin{split}\nonumber 
\norm{a_{\mu}(f)a_{\lambda}(f)\psi}&\leq
\left( \int \frac{\va{\rho(k)}^2}{\vak^{2}}d^3k\right)
\norm{(H_{ph} +1)\psi}\\  &+K\left( \int 
\frac{\va{\rho(k)}^2}{\vak^{2}}d^3k\right)^{1/2}    
\left(\int {\va{\rho(k)}^2}d^3k\right)^{1/2}       
\norm{(H_{ph} +1)^{1/2}\psi}\ ,\\
\norm{a^*_{\mu}(f)a_{\lambda}(f)\psi}&\leq
\left( \int 
\frac{\va{\rho(k)}^2}{\vak^{2}}d^3k\right)
\norm{(H_{ph} +1)\psi}\\
&+\left( K\left( \int 
\frac{\va{\rho(k)}^2}{\vak^{2}}d^3k\right)^{1/2}    
\left( \int {\va{\rho(k)}^2}d^3k\right)^{1/2}\right.\\
&\left. +\left(\int \frac{\va{\rho(k)}^2}{\vak^{2}}d^3k\right)^{1/2}
\left(\int \frac{\va{\rho(k)}^2}{\vak}d^3k\right)^{1/2}\right)
\norm{(H_{ph} +1)^{1/2}\psi}\ ,
\\\norm{a^*_{\mu}(f)a^*_{\lambda}(f)\psi}&\leq
\left( \int 
\frac{\va{\rho(k)}^2}{\vak^{2}}d^3k\right)
\norm{(H_{ph} +1)\psi}\\
&+\left( K\left( \int 
\frac{\va{\rho(k)}^2}{\vak^{2}}d^3k\right)^{1/2}    
\left( \int {\va{\rho(k)}^2}d^3k\right)^{1/2}\right.\\
&\left. +\left(\int \frac{\va{\rho(k)}^2}{\vak^{2}}d^3k\right)^{1/2}
\left(\int \frac{\va{\rho(k)}^2}{\vak}d^3k\right)^{1/2}\right)
\norm{(H_{ph} +1)^{1/2}\psi}\\
&+\left( \left( \int 
\frac{\va{\rho(k)}^2}{\vak^{2}}d^3k\right)^{1/2}    
\left(\int \vak {\va{\rho(k)}^2}d^3k\right)^{1/2}       
+\int \frac{\va{\rho(k)}^2}{\vak}d^3k\right)\norm{\psi}
\end{split}\end{align}     
where $K=\frac{1}{\pi}\int_{0}^\infty \frac{\sqrt 
\lambda}{(1+\lambda)^2}$.

As $(H_{ph}+1)^{1/2}$ is relatively bounded with respect to 
$H_{ph}+1$ (and thus to $H_{0}$) with a zero relative bound,
we deduce that the relative bound of $\frac{g^2}{2m}
     :A(x,\rho)\cdot A(x,\rho):$ with respect to $H_{0}$ is given by
$$16\frac{g^2}{2m}\frac{1}{4\pi^2}
 \int \frac{\va{\rho(k)}^2}{\vak^{2}}d^3k\ .$$
Finally we get that $H_{I}$ is relatively bounded with respect to 
$H_{0}$ with the relative bound
$$\frac{12\va{g}}{\pi \sqrt{2m}}\left(\int \frac{\va{\rho(k)}^{2}}{\vak^2}d^3k \right)^{1/2} +
\frac{2g^2}{\pi^2m}\int \frac{\va{\rho(k)}^{2}}{\vak^2}d^3k$$
and hence theorem \ref{thm2} is a consequence of the Kato-Rellich theorem.
\qed

\subsection{The reduced Hamiltonian}\label{reduc}
The operator $H$ is invariant by translation in the $x_{3}$-direction. 
Thus, denoting by $P_{3}$ the total momentum in the $x_{3}$-direction
($P_{3}= p_{3}\otimes 1 + 1 \otimes {\rm d}\Gamma (k_{3})$), $H$ has a direct integral representation in a spectral 
representation of $P_{3}$, i.e. 
\be \label{Hrep} 
H \simeq \int_{\R}^{\oplus}H(P_{3}) dP_{3}
\ee
To compute $H(P_{3})$ we proceed as in \cite{F74}  (see also 
\cite{AHS,A00} and\cite{FGScom}). Let $\Pi$ be the unitary map from $\mathcal 
H$ to 
$L^2(\R)\otimes L^2(\R^2,\C^2)\otimes \mathcal F$ defined by
$$
(\Pi \phi)_{n}(P_{3},x',k_{1},\ldots,k_{n})=\hat 
\phi_{n}(x',P_{3}-\sum_{i=1}^n k_{i,3},x',k_{1},\ldots,k_{n})
$$
where
the hat stands for the partial Fourier transform in $x_{3}$.
One easily verifies that, on $C^\infty_{0}(\R^3,\C^2)\otimes \mathcal F_{0,fin}$,
$$
\Pi A_{j}(x',x_{3},\rho)\Pi^*=A_{j}(x',0,\rho)\ .
$$
Therefore, for $\psi \in C^\infty_{0}(\R^3,\C^2)\otimes \mathcal F_{0,fin}$,
$$
(\Pi H\Pi^*\psi)(P_{3},\cdot)=H(P_{3})\psi(P_{3},\cdot)
$$
where the reduced Hamiltonian $H(P_{3})$ is given by
\be \label{HP3}
H(P_{3})= H_{0}(P_{3}) + H_{I}(P_{3})
\ee
with
\begin{eqnarray}\label{H0P3}
    H_{0}(P_{3})=h(b,V) \otimes 1 + 1 \otimes 
   \left\{\frac{1}{2m}(P_{3}-{\rm d}\Gamma (k_{3}))^2+ H_{ph}\right\} 
\end{eqnarray}
and
\begin{align}\begin{split}\label{HIP3}
    H_{I}&(P_{3})=-\frac{g}{2m}\sigma \cdot B(x',0,\rho)
    -\frac{g}{m}\sum_{j=1,2}
    A_{j}(x',0,\rho)(p_{j}-ea_{j}(x')) \\
     &-\frac{g}{m }A_{3}(x',0,\rho)(P_{3}-{\rm d}\Gamma (k_{3}))
     +\frac{g^2}{2m}
     :A(x',0,\rho)\cdot A(x',0,\rho):
\end{split}\end{align}
For every $P_{3}$, $H(P_{3})$ is now an operator in 
$L^2(\R^2,\C^2)\otimes \mathcal F$. We want to show that this formal operator 
defines a self-adjoint one such that \eqref{Hrep} is satisfied.

The operator $\frac{1}{2m}(P_{3}-{\rm d}\Gamma (k_{3}))^2+ H_{ph}$ is 
essentially self-adjoint on $\mathcal F_{0,fin}$. Therefore, for 
every $P_{3}\in \R$,
$H_{0}(P_{3})$ is essentially self-adjoint in
$C^\infty_{0}(\R^2,\C^2)\otimes \mathcal F_{0,fin}$. We still denote 
by $H_{0}(P_{3})$ its self-adjoint extension. On the other hand 
$H_{I}(P_{3})$ is a symmetric operator on 
$C^\infty_{0}(\R^2,\C^2)\otimes \mathcal F_{0,fin}$ and we want to 
prove that it is relatively bounded with respect to $H_{0}(P_{3})$. 
We then follow closely the lines of section \ref{sa} and we only 
focus on the estimates of the new terms.
For $\psi \in C^\infty_{0}(\R^2,\C^2)\otimes \mathcal F_{0,fin}$ we 
have
\begin{align}\begin{split}\label{A3}\nonumber
\frac{\vag}{m}\norm{A_{3}(x',0,\rho)(P_{3}-d\Gamma (k_{3}))\psi}&\leq
\frac{4\vag}{\pi\sqrt{2m}}\left( \int 
\frac{\va{\rho(k)}^2}{\vak^2}d^3k\right)^{1/2}\\ &\norm{ (I\otimes 
H_{ph}^{1/2})(I\otimes \frac{P_{3}-d\Gamma (k_{3})}{\sqrt{2m}})\psi}\\
+\frac{2\vag}{\pi\sqrt{2m}}\left( \int 
\frac{\va{\rho(k)}^2}{\vak}d^3k\right)^{1/2}& \norm{ 
(I\otimes \frac{P_{3}-d\Gamma (k_{3})}{\sqrt{2m}})\psi}\ .
\end{split}\end{align} 

For every component $\psi_{n}$ of $\psi\in \mathcal F_{0,fin}$ 
associated with $n$ photons the operator 
$(I\otimes 
H_{ph}^{1/2})(I\otimes \frac{P_{3}-d\Gamma (k_{3})}{\sqrt{2m}})$ is 
the multiplication operator by the function 
$(\sum_{i=1}^n \omega(k_{i}))^{1/2}\frac{P_{3}-\sum_{i=1}^n 
k_{i,3}}{\sqrt{2m}}$. We then get
\begin{align}\begin{split}\nonumber
\norm{ (I\otimes 
H_{ph}^{1/2})(I\otimes \frac{P_{3}-d\Gamma 
(k_{3})}{\sqrt{2m}})\psi}&\leq \frac{1}{\sqrt 2}
\norm{ (I\otimes \left\{
H_{ph}+\frac{1}{{2m}}(P_{3}-d\Gamma 
(k_{3}))^2\right\})\psi}\\
&\leq \frac{1}{\sqrt 2}
\norm{(H_{0}(P_{3})-e(b,V))\psi}
\end{split}\end{align} 
and, for any $\epsilon >0$,
$$
\norm{(I\otimes \frac{P_{3}-d\Gamma 
(k_{3})}{\sqrt{2m}})\psi}\leq \sqrt{\frac{\epsilon}{2}}
\norm{(H_{0}(P_{3})-e(b,V))\psi} + 
\frac{1}{\sqrt{2\epsilon}}\norm{\psi}\ .
$$
Therefore
\begin{align}\begin{split}\nonumber
\frac{\vag}{m}\|A_{3}(x',0,\rho)&
(P_{3}-d\Gamma (k_{3}))\psi \| \leq
\frac{2\vag}{\pi\sqrt{m}}\left( \int 
\frac{\va{\rho(k)}^2}{\vak^2}d^3k\right)^{1/2}
\norm{(H_{0}(P_{3})-e(b,V))\psi} \\
&+\sqrt{\frac{\epsilon}{m}}\frac{\vag}{\pi}\left( \int 
\frac{\va{\rho(k)}^2}{\vak}d^3k\right)^{1/2}\norm{(H_{0}(P_{3})-e(b,V))\psi}\\
&+\frac{1}{\sqrt{\epsilon m}}\frac{\vag}{\pi}\left( \int 
\frac{\va{\rho(k)}^2}{\vak}d^3k\right)^{1/2}\norm{\psi}
\ .
\end{split}\end{align}
Thus, as in section \ref{sa}, we obtain that, for any $\eta >0$, 
there exists a finite constant $a_{\eta}$ such that
\be \label{estimHI}
\norm{H_{I}(\pp)\psi}\leq \vag
(b+\eta)\norm{\hop\psi}+\vag a_{\eta}\norm\psi
\ee
with 
$$
b=\frac{12}{\pi \sqrt{2m}}\left(\int \frac{\va{\rho(k)}^{2}}
{\vak^2}d^3k \right)^{1/2} +
\frac{2g}{\pi^2m}\int \frac{\va{\rho(k)}^{2}}{\vak^2}d^3k\ .
$$
Therefore we have
\begin{theorem}\label{thm3}
Assume \eqref{rho}
and
\be \label{rhosa}
\frac{6\va{g}}{\pi \sqrt{2m}}\left(\int \frac{\va{\rho(k)}^{2}}{\vak^2}d^3k \right)^{1/2} +
\frac{g^2}{\pi^2m}\int \frac{\va{\rho(k)}^{2}}{\vak^2}d^3k    <\frac{1}{2}\ .
\ee    
Then, for every $P_{3}\in \R$, $H(P_{3})$ is a self-adjoint operator in 
$L^2(\R^2,\C^2)\otimes \mathcal F$ with domain 
$D(H(P_{3}))=D(H_{0}(P_{3}))$ and $H(P_{3})$ is essentially self-adjoint on
$C^\infty_{0}(\R^2,\C^2)\otimes \mathcal F_{0,fin}$.
\end{theorem} 
Further we get 
\begin{corollary}\label{decomp}
    We have
    $$
    \Pi H \Pi^*=\int_{\R}^\oplus H(P_{3})dP_{3}\ .
    $$
    \end{corollary}
     
The proof of corollary \ref{decomp} follows by mimicking \cite{A00}.

\section{Main results}
For a bounded below self-adjoint operator $A$ with a ground state, 
$m(A)$ will denote the multiplicity of $\inf \sigma (A)$. Our main 
result is the following theorem which states that, for $P_{3}$ and 
$g$ sufficiently small, $H(P_{3})$ has a ground state:
\begin{theorem}\label{mainthm}
Assume that the cutoff  function satisfies \eqref{rho}, \eqref{rhosa} 
and
\be\label{rho2}
    \int_{\vak \leq 1} \frac{\va{\rho(k)}^{2}}{\vak^{3}}d^3k <\infty \ .
    \ee
Then there exist $P>0$ and $g_{0}>0$ such that for $|P_{3}|\leq P$ 
and $\vag \leq g_{0}$, $H(P_{3})$ has a ground state. Furthermore $ 
m(H(P_{3}))\leq m(h(b,V))$. In particular, if $e(b,V)$ is a simple 
eigenvalue of $h(b,V)$, then $\inf \sigma (H(P_{3}))$ is a simple 
eigenvalue too.
\end{theorem} 
The proof of this theorem is given in the next section.

\begin{remark}\ \newline\label{r7} Notice that
the regularization condition \eqref{rho2} does not 
allow $\rho (k)=1$ near the origin. According to \cite{Chen}, one may 
conjecture that $H(P_{3})$ has no ground 
state for $P_{3}\neq 0$ when this infrared condition is not 
satisfied.\end{remark}

The existence of a ground state has several consequences. The first 
one is the existence of asymptotic Fock representations for the CCR.

For $f\in L^2(\R^3,\C^2)$, we define on $D(H_{0}(P_{3}))$ the 
operators 
$$
a^\sharp_{\mu,t}(f):= e^{itH(P_{3})}e^{-itH_{0}(P_{3})}a^\sharp_{\mu}(f)
e^{itH_{0}(P_{3})}e^{-itH(P_{3})}\ .
$$
Let $Q$ be a closed null set such that the polarization vectors 
$\epsilon_{\mu}(k)$ are $C^\infty$ on $\R^3\setminus Q$ for $\mu =1,2$.
We have
\begin{corollary}
Suppose that the hypothesis of theorem \ref{mainthm} are satisfied.
Then, for $f\in C_{0}^\infty(\R^3\setminus Q)$ and for every $\Psi 
\in D(H_{0}(P_{3}))$ the strong limits of $ a^\sharp_{\mu,t}(f)$ exist:
$$
\lim_{t\to \pm \infty}a^\sharp_{\mu,t}(f)\Psi 
=:a^\sharp_{\mu,\pm}(f)\Psi \ .
$$
The $a^\sharp_{\mu,\pm}$'s satisfy the CCR and, if $\Phi (P_{3})$ is 
a ground state for $H(P_{3})$, we have for $f\in C_{0}^\infty(\R^3\setminus Q)$
and $\mu=1,2$
$$
a_{\mu,\pm}(f)\Phi(P_{3})=0 \ .
$$    
\end{corollary}

We then deduce the following corollary
\begin{corollary}
    Under the hypothesis of theorem \ref{mainthm},
    the absolutely 
    continuous  spectrum of $H(P_{3})$  equals to $[\inf 
    \sigma (H(P_{3})), +\infty)$.
    \end{corollary}

The proofs of these two corollaries follow by mimicking \cite{H01,H04}.

Our last application concerns the renormalized mass  
and magnetic moment of the electron. 

From now on we assume that 
 the ground state of $h(b,V)$ is simple in such a 
way that the ground state of $H(P_{3})$ is also simple.
To state our result we need the 
notations and results of theorem \ref{thm6}. Let $\esp$ 
be the ground energy of the hamiltonian with infrared cutoff $\hsp$, 
$\esp$ is a simple and an isolated eigenvalue of $\hsp$ and therefore we 
deduce from the standard Kato-Rellich perturbation theory that $\esp$ is a 
regular function of $P_{3}$. We then define the renormalized mass of 
the dressed electron by 
\be \label{mstar}
m^\star:=\liminf_{\sigma \to 0} m^\star_{\sigma}
\ee
where
$$
(m^\star_{\sigma})^{-1}=\partial^2_{\pp} E_{\sigma} (0) \ .
$$
Let ${g}_{\mathrm{el}}$ be the magnetic moment of the dressed electron. 
We then have
\begin{corollary}\label{corom}
    Under the hypothesis of theorem \ref{mainthm} and assuming that
    $e(b,V)$ is a simple eigenvalue, we have 
    $$m^\star \geq m$$
    i.e. the renormalized mass of the dressed electron in a magnetic 
    field is larger than or equal to the bare mass of the electron. It then 
    follows that ${g}_{\mathrm{el}}\geq 2$.
\end{corollary}

\proof
Since the ground state of $\hsp$ is non degenarate, $\esp$ and $\psp$ are smooth 
function of the parameter $\pp$ and we easily obtain by differentiating 
the relation
$\hsp \psp = \esp \psp$ the following formulas
$$\partial_{\pp} \esp =\langle \psp, (\partial_{\pp} \hsp )\psp \rangle
$$
and
\begin{align}\nonumber
    \partial^2_{\pp} \esp =&\langle \psp, (\partial^2_{\pp} \hsp )\psp 
    \rangle\\
-&2\langle \partial_{\pp} \psp , (\hsp -\esp )\partial_{\pp} \psp 
\rangle \ .\end{align}
As $\partial^2_{\pp} \hsp = 1/m$ and $\hsp -\esp\geq 0$
we obtain $m^\star_{\sigma}\geq m$ for all $\sigma$ and thus $m^\star \geq m$.

The fact that $\mathbf{g}_{\mathrm{el}}\geq 2$ follows from $m^\star 
\geq m$ as in \cite{Coh84,Coh96,Sp04}.
\qed

\section{Proof of the main theorem}
%\subsection{sketch of the proof}
To begin with we introduce an infrared regularized cutoff in the interaction 
Hamiltonian $H_{I}(P_{3})$. Precisely, for $\sigma >0$, let 
$\rho_{\sigma}$ be a $C_{0}^\infty$ regularization of $\rho$ such that
\begin{itemize}
    \item[(i)] $\rho_{\sigma}(k)=0$ for $\vak\leq \sigma$
    %\item[(ii)] $\lim_{\sigma \to 0}\rho_{\sigma}(k)=\rho(k)$ a.e. 
    %$k\in \R^3$.
    \item[(ii)] $\lim_{\sigma \to 0}\int 
    \frac{\va{\rho_{\sigma}(k)-\rho(k)}^2}{\vak^j}d^3k=0$ for $j=1,2,-1$.
\end{itemize}
We define $H_{I,\sigma}(P_{3})$ as the operator obtained from 
\eqref{HIP3} by substituting $\rho_{\sigma}(k)$ for $\rho(k)$. We then 
introduce
$$
H_{\sigma}(P_{3})=H_{0}(P_{3})+H_{I,\sigma}(P_{3})
$$
and we set $E_{\sigma}(P_{3}):=\inf \sigma (H_{\sigma}(P_{3}))$.
Theorem \ref{mainthm} is a simple consequence of the following 
result (see \cite{BFS98b})
\begin{theorem}\label{thm6}
There exist $g_{0}>0$, $\sigma_{0}>0$ and $P>0$ such that, for every $g$ satisfying 
$\vag\leq g_{0}$, for every $\sigma$ satisfying $0<\sigma <\sigma_{0}$ 
and for every $P_{3}$ satisfying $\va{P_{3}}\leq P$, 
the following properties hold:
\begin{itemize}
\item[(i)] For every $\Psi \in D(H_{0}(P_{3}))$ we have 
$H_{\sigma}(P_{3})\Psi \to_{\sigma \to 0}H(P_{3})\Psi$
\item[(ii)]  $H_{\sigma}(P_{3})$ has a 
normalized ground state $\Phi_{\sigma}(P_{3})$.
\item[(iii)] Fix $\lambda \in (e(b,V),0)$. We 
have
$$
\langle \Phi_{\sigma}(P_{3}),P_{(-\infty,\lambda]}\otimes 
P_{\Omega_{ph}} \ \Phi_{\sigma}(P_{3})\rangle \geq 
1-\delta_{g}(\lambda)$$
where $\delta_{g}(\lambda)$ tends to zero when $g$ tends to zero 
and $\delta_{g}(\lambda)<1$ for $\vag\leq g_{0}$.
\end{itemize}

\end{theorem}
In the last item, $P_{(-\infty,\lambda]}$ is the spectral projection on
$(-\infty,\lambda]$ associated to $h(b,V)$ and 
$P_{\Omega_{ph}}$ is the orthogonal projection on $\Omega_{ph}$, the 
vacuum state in $\mathcal F$.

Theorem \ref{mainthm} is easily deduced from theorem \ref{thm6} as 
follows. Let $\Phi_{\sigma}(P_{3})$ be as in theorem \ref{thm6} (ii). Since 
$\norm{\Phi_{\sigma}(P_{3})}=1$, there exits a sequence $(\sigma_{k})_{k\geq 
1}$ converging to zero such that $(\Phi_{\sigma_{k}}(P_{3}))_{k\geq1}$ 
converges weakly to a state $\Phi(P_{3})$. On the other hand, since 
$P_{(-\infty,\lambda]}\otimes 
P_{\Omega_{ph}}$ is finite rank for $\lambda \in (e(b,V),0)$, it 
follows from (iii) that for $\vag\leq g_{0}$ and $|P_{3}|\leq P_{0}$,
$$
\langle \Phi(P_{3}),P_{(-\infty,\lambda]}\otimes 
P_{\Omega_{ph}} \ \Phi(P_{3})\rangle \geq 
1-\delta_{g}(\lambda)$$
which implies $\Phi(P_{3})\neq 0$. Then we deduce from (i) and from a 
well known result (\cite{AH97} lemma 4.9) that $\Phi(P_{3})$ 
is a ground state for $H(P_{3})$. 

The result concerning the multiplicity of the ground state is an easy 
consequence of corollary 3.4 in \cite{H04}.

So it remains to prove theorem \ref{thm6}. The assertion (i) is 
easily verified in section \ref{i} below. The second assertion is proved 
in the appendix. Actually the proof of (ii) is lenghty but straightforward 
since with the infrared cutoff we have a control of the photon's 
number in term of the energy. The real difficult part is the third 
one which allows to  relax the infrared cutoff. 
%Actually, assertion 
%(iii) states that the electronic part of a ground state must have 
%a contribution on the eigenfunctions of $h(b,V)$  of negative energy and 
%that this 
%contribution is controlled independently of $\sigma$.   
 The fundamental 
lemma in the proof of (iii) is  lemma \ref{lem8} which states that, 
for $g$ and $P_{3}$ small enough, 
the difference $E_{\sigma}(P_{3}-k_{3})-E_{\sigma}(P_{3})$
is minorized by $-\frac{3}{4}\vak$ uniformly with respect to $\sigma$. 
This estimate, proved in section \ref{fonda},
is essential to control the  number of photons in a ground state of 
$H_{\sigma}(P_{3})$ via a pull 
through formula (cf. section \ref{iii}).

\subsection{Proof of (i) of theorem \ref{thm6}}\label{i}
Let $\tilde\rho_\sigma:=\rho -\rho_\sigma$. We have
\begin{align}\begin{split}\nonumber
    H(P_{3})-&H_{\sigma}(P_{3})=H_{I}(P_{3})-H_{I,\sigma}(P_{3})\\
&=-\frac{g}{2m}\sigma \cdot B(x',0,\tilde\rho_\sigma)
    -\frac{g}{m}\sum_{j=1,2}
    A_{j}(x',0,\tilde\rho_\sigma)(p_{j}-ea_{j}(x')) \\
     &-\frac{g}{m }A_{3}(x',0,\tilde\rho_\sigma)(P_{3}-{\rm d}\Gamma (k_{3}))
     +\frac{g^2}{2m}
     :A(x',0,\tilde\rho_\sigma)\cdot A(x',0,\rho): \\
     &+\frac{g^2}{2m}
     :A(x',0,\rho_\sigma)\cdot A(x',0,\tilde\rho_\sigma):
\end{split}\end{align}
Therefore, as by hypothesis $\lim_{\sigma \to 0}\int 
    \frac{\va{\tilde\rho_{\sigma}(k)}^2}{\vak^j}d^3k=0$ for 
    $j=-1,1,2$, we deduce from the estimates of sections \ref{sa}, 
    \ref{reduc} 
    and from the Lebesgue's theorem that for every $\Psi \in 
    D(H_{0}(P_{3}))$,
    $$
   (H(P_{3})- H_{\sigma}(P_{3}))\Psi \to_{\sigma \to 0}0\ .
    $$
    \qed
\subsection{Fundamental estimates}\label{fonda}
In this section we give two lemmas which allow to control the 
function$$\pp \mapsto \esp\ .$$
Let $g_{1}>0$ such that \eqref{rhosa} is satisfied for $\vag\leq 
g_{1}$.
\begin{lemma}\label{lem7}
There exist $\sigma_{0}>0$ and a finite constant $C>0$ which does not depend on 
$\sigma \in (0,\sigma_{0}]$ and $\pp \in \R$ such that
\be \label{estime}
e(b,V)-\vag C\leq \esp \leq e(b,V) +\frac{\pp^2}{2m}    
\ee
for every $\sigma \in (0,\sigma_{0}]$, $\pp \in \R$ and $\vag\leq g_{1}$.
\end{lemma}
\proof
Let $\phi(b,V)$ be the normalized ground state of $h(b,V)$. Since 
$\langle a_{\mu}(k)\Omega_{ph},\om \rangle = \langle \om , 
a^*_{\mu}(k)\om \rangle = 0$, we have
\begin{align}\begin{split}\nonumber
\langle \hsp \phi(b,V)\otimes \om ,\phi(b,V)\otimes \om\rangle &=
\langle \hop \phi(b,V)\otimes \om ,\phi(b,V)\otimes \om\rangle\\
&=e(b,V) +\frac{\pp^2}{2m}
\end{split}\end{align}
and thus
\begin{align}\begin{split}\nonumber
\esp&:=\inf \left\{\langle \hsp \Phi ,\Phi\rangle  \mid \Phi \in D(H_{0}(\pp)),\ 
\norm\Phi =1\right\}\\
&\leq e(b,V) +\frac{\pp^2}{2m}\ .
\end{split}\end{align}
On the other hand, 
let $ \ths$ be the following operator in 
$L^2(\R^2,\C^2)\otimes \F$:
$$
\ths = \tho +\this
$$
with
$$
\tho= h(b,V)\otimes I + I\otimes H_{ph}
$$
and
\begin{align}\begin{split}\nonumber
    \this &=-\frac{g}{2m}\sigma \cdot B(x',0,\rho_{\sigma})\\
    &-\frac{g}{2m}\sum_{j=1,2}
    \left(A_{j}(x',0,\rho_{\sigma})(p_{j}-ea_{j}(x')) 
    +(p_{j}-ea_{j}(x'))A_{j}(x',0,\rho_{\sigma})\right)\\
     &+\frac{g^2}{2m}\sum_{j=1,2}
     :A_{j}(x',0,\rho_{\sigma})\cdot A_{j}(x',0,\rho_{\sigma}):
\end{split}\end{align} %

As in section \ref{reduc} one easily checks that, for $\vag \leq 
g_{1}$, $\ths$ is a self-adjoint operator in $L^2(\R^2,\C^2)\otimes \F$ 
with domain $D(H_{0}(\pp))$. Furthermore, on $D(H_{0}(\pp))$
$$
\hsp =\ths 
+\frac{1}{2m}(P_{3}-d\Gamma(k_{3})-gA_{3}(x',0,\rho_{\sigma}))^2 - 
C(g,\sigma)
$$
where, in order to take into account the Wick normal ordering,
$$
C(g,\sigma):=\frac{g^2}{2m}\frac{1}{(2\pi)^2}\int 
\frac{|\rho_{\sigma}|^2}{\vak}\left(\sum_{\mu 
=1,2}\epsilon_{\mu}(k)_{3}^2\right)d^3k \ .
$$
Hence,
\be\label{tilde1}\inf \sigma (\ths)\leq \esp + C(g,\sigma)
\ee
for every $\pp\in\R$. Furthermore there exists $\sigma_{0}>0$ such 
that for $0<\sigma\leq \sigma_{0}$ 
$$
C(g,\sigma) \leq \tilde C g^2
$$
whith
$$
\tilde C = \frac{1}{m}\frac{1}{(2\pi)^2}\left( \int 
\frac{|\rho|^2}{\vak}d^3k +1\right) \ .
$$
By \eqref{estimHI} which also holds when $\rho$ is replaced by 
$\rho_{\sigma}$, we get that there exist two constants $b>0$, 
$a>0$ which do not depend on $\sigma\in(0,\sigma_{0}]$ and  $g \in [- 
g_{1},g_{1}]$ and satisfying $bg_{1}<1$ such that for $\Phi\in D(\tho)$ 
and for $\sigma\in(0,\sigma_{0}]$ with $\sigma_{0}$ sufficiently small
$$
\norm{\this \Phi}\leq \vag (b\norm{\tho \Phi}+a\norm\Phi)\ .
$$
Therefore, since $\inf \sigma(\tho)=e(b,V)$, we obtain as a consequence of the Kato-Rellich theorem,
\be \label{tilde2}
\inf \sigma (\ths)\geq e(b,V) -\max 
\left(\frac{a\vag}{1-b\vag}\ ,\ a\vag+b\vag |e(b,V)|\right)\ .
\ee
Combining \eqref{tilde1} and \eqref{tilde2} we deduce the announced lower bound 
for $\esp$ with 
$$
C=\max 
\left(\frac{a}{1-bg_{1}}\ ,\ a+b |e(b,V)|\right)+\frac{g_{1}}{m}\frac{1}{(2\pi)^2}\left( \int 
\frac{|\rho|^2}{\vak}d^3k +1\right) \ .
$$
\qed

\begin{lemma}\label{lem8}
There exist $0<g_{2}\leq g_{1}$ and $\alpha>0$ such that
\be\label{estimee}
E_{\sigma}(\pp -k_{3})-\esp \geq -\frac{3}{4}\vak
\ee
uniformly for $k\in \R^3$, $\sigma\in(0,\sigma_0 ]$, $\vag\leq g_{2}$ and 
$|P_{3}|\leq \alpha$.    
\end{lemma}
Remark that in this lemma we do not assume that $\hsp$ has a ground 
state (i.e. we do not assume that $\esp$ is an eigenvalue of $\hsp$)
and actually we will use \eqref{estimee} in appendix \ref{Ap} 
where we prove the existence of a ground state for the Hamiltonian with 
infrared cutoff.
\proof
First we remark that, if \eqref{estimee} is proved for $\hsp +c$ for 
some constant $c$, it will hold also for $\hsp$. Thus, in what 
follows, we suppose that $e(b,V)=0$.

The proof decomposes in two steps. In the first one, we consider the 
large values of $\va{k_{3}}$ (namely $\vaka \geq m/4$) while, in the 
second one, we consider the small values of $\vaka$ (namely $\vaka 
\leq m/4)$. 

From \eqref{estime}, we deduce that, uniformly for $\sigma\in(0,\sigma_0 ]$ 
and $\vag\leq g_{1}$, we have for all $k$ and $P_{3}$
$$
E_{\sigma}(P_{3}-k_{3})-\esp \geq -\frac{\pp^2}{2m} -C\vag
$$
and thus assuming $|P_{3}|\leq {\frac{\sqrt 3}{4}}m$ and $\vag \leq 
\frac{3m}{32C}$, \eqref{estimee} holds true for $\vaka \geq m/4$.

\smallskip

Now we suppose $\vaka \leq m/2$.
As $E_{\sigma}(P_{3}-k_{3})$ belongs to the spectrum of $H_{\sigma}(P_{3}-k_{3})$
there exists a sequence $(\psi_{j})_{j\geq 1}$ in 
$D(H_{\sigma}(P_{3}-k_{3}))$ ($=D(H_{0}(P_{3}=0))$) such that 
$\norm{\psi_{j}}=1$ and 
$$\lim_{j\to \infty} H_{\sigma}(P_{3}-k_{3})\psi_{j}-E_{\sigma}(P_{3}-k_{3})\psi_{j} 
= 0\ .$$
We then have for every $j$
\begin{align}\begin{split}\label{A1}
\langle H_{\sigma}(P_{3}-k_{3})\psi_{j} , \psi_{j}\rangle &=
\langle H_{\sigma}(P_{3})\psi_{j} , \psi_{j}\rangle +\frac{k_{3}^2}{2m}
-\frac{k_{3}}{m}\langle (P_{3}-d\Gamma(k_{3}))\psi_{j} , 
\psi_{j}\rangle \\
&+\frac{2gk_{3}}{m}\langle A_{3}(x',0,\rho_{\sigma})\psi_{j} , 
\psi_{j}\rangle\\
&\geq \esp +\frac{k_{3}^2}{2m} -
\frac{|k_{3}|}{m}|\langle (P_{3}-d\Gamma(k_{3}))\psi_{j} , 
\psi_{j}\rangle |\\
&-\frac{2\vag |k_{3}|}{m}|\langle A_{3}(x',0,\rho_{\sigma})\psi_{j} , 
\psi_{j}\rangle | \ .
\end{split}\end{align}
In what follows $C$ will denote any positive constant which does not 
depend on $\pp\in \R$, $k_{3}\in \R$, $g\in[-g_{1},g_{1}]$, $\sigma \in (0,\sigma_0 ]$ 
and $j\geq 1$. We have
\begin{align}\begin{split}\label{A2}
    |\langle (P_{3}-d\Gamma(k_{3}))\psi_{j} , 
    \psi_{j}\rangle |&\leq \vaka +|\langle (P_{3}-k_{3}-d\Gamma(k_{3}))\psi_{j} , 
\psi_{j}\rangle |\\
&\leq \vaka + \norm{(P_{3}-k_{3}-d\Gamma(k_{3}))\psi_{j}}\\
&\leq \vaka + \norm{(P_{3}-k_{3}-d\Gamma(k_{3}))^2\psi_{j}}^{1/2}\\
&\leq \vaka +\sqrt{2m} \norm{H_{0}(P_{3}-k_{3})\psi_{j}}^{1/2}\ .
\end{split}\end{align}
On the other hand, we get from \eqref{aa} and \eqref{aaa},
\begin{align}\begin{split}\label{A3}
    |\langle A_{3}(x',0,\rho_{\sigma}\psi_{j} , 
    \psi_{j}\rangle |&\leq C(\norm{H_{ph}^{1/2}\psi_{j}}+1)\\
    &\leq C (\norm{H_{0}(P_{3}-k_{3})\psi_{j}}^{1/2}+1)\ .
\end{split}\end{align}
Now, given $\epsilon>0$, let $J$ be such that
\be\label{A5}
\norm{H_{\sigma}(P_{3}-k_{3})\psi_{j}-E_{\sigma}(P_{3}-k_{3})\psi_{j}}\leq 
\epsilon
\ee
for every $j\geq J$ (notice that $J$ depends on $\epsilon$ but also 
on $\sigma$ and $\pp-k_{3}$).

Inserting \eqref{A2} and \eqref{A3} in \eqref{A1} we obtain for 
$j\geq J$
\begin{align}\begin{split}\label{A4}
    E_{\sigma}(P_{3}-k_{3})-\esp &\geq -\epsilon  -\frac{k_{3}^2}{2m}
    -\vaka \sqrt{2/m}\norm{H_{0}(P_{3}-k_{3})\psi_{j}}^{1/2}\\
    &-\vaka C\vag(\norm{H_{0}(P_{3}-k_{3})\psi_{j}}^{1/2}+1)
\end{split}\end{align}
and it remains to estimate $\norm{H_{0}(P_{3}-k_{3})\psi_{j}}$.

Writing 
$$
H_{0}(P_{3}-k_{3})\psi_{j}=(H_{\sigma}(P_{3}-k_{3}) -E_{\sigma}(P_{3}-k_{3}))
\psi_{j} + E_{\sigma}(P_{3}-k_{3})\psi_{j} -H_{I,\sigma}(P_{3}-k_{3}) 
\psi_{j}$$
we get for $j\geq J$
$$\norm{H_{0}(P_{3}-k_{3})\psi_{j}}\leq \epsilon
+|E_{\sigma}(P_{3}-k_{3})|+\norm{H_{I,\sigma}(P_{3}-k_{3}) 
\psi_{j}}\ .$$
Using \eqref{estimHI} there exists $C>0$ such that
$$\norm{H_{I,\sigma}(P_{3}) 
\phi}\leq \vag C(\norm{H_{0}(P_{3}) 
\phi}+1)$$ for every $P_{3}\in \R$ and $\phi \in D(H_{0}(\pp))$, 
$||\phi ||\leq 1$.
Thus, choosing $g'_{1}\leq g_{1}$ such that $g'_{1} C\leq 
1/2$, we get
\be \label{estimH0}
\norm{H_{0}(P_{3}-k_{3}) 
\psi_{j}}\leq 2\epsilon + 2 |E_{\sigma}(P_{3}-k_{3})| + 2\vag C
\ee
for $j\geq J$ and $\vag\leq g'_{1}$. 
Inserting this last inequality in \eqref{A4} we obtain
\begin{align}\begin{split}\nonumber
    E_{\sigma}(P_{3}-k_{3})-\esp &\geq -\epsilon  -\frac{k_{3}^2}{2m}
    -2\vaka \sqrt{1/m}(\epsilon +  |E_{\sigma}(P_{3}-k_{3})| + \vag C)^{1/2}\\
    &-\vaka C\vag((2\epsilon + 2 |E_{\sigma}(P_{3}-k_{3})| + 2\vag C)^{1/2}+1)
\end{split}\end{align}
for every $\epsilon >0$. Hence
\begin{align}\begin{split}\label{A6}
    E_{\sigma}(P_{3}-k_{3})-\esp &\geq   -\vaka\left\{\frac{\vaka}{2m}
    +2 \sqrt{1/m}(   |E_{\sigma}(P_{3}-k_{3})| + \vag C)^{1/2}\right.\\
    &\left.+ C\vag(( 2 |E_{\sigma}(P_{3}-k_{3})| + 2\vag 
    C)^{1/2}+1)\right\}
\end{split}\end{align}
for every $k_{3}$ and $P_{3}$ in $\R$.
Finally we use \eqref{estime} to 
get for $|k_{3}|\leq m/4$,
$$|E_{\sigma}(P_{3}-k_{3})|\leq C\vag +\frac{\pp^2}{2m} 
+\frac{|\pp|}{4}+\frac{m}{32}$$ and therefore there exit $\alpha>0$ 
and $g_{2}\leq g'_{1}$ such that for $|\pp|\leq \alpha$, $|k_{3}|\leq 
m/4$ and $\vag\leq g_{2}$,
\begin{align}\begin{split}\nonumber
    E_{\sigma}(P_{3}-k_{3})-\esp &\geq -\frac{3}{4}|k_{3}|\ .
\end{split}\end{align}

\qed
\subsection{Proof of (iii) of theorem \ref{thm6}}\label{iii}

In this section we assume that assertion (ii) of theorem \ref{thm6} is 
already proved (see appendix \ref{Ap}). Thus let $\psp$ 
denote a normalized ground state of $\hsp$, i.e.
$$
\hsp \psp = \esp \psp\ .
$$
The main problem in proving (iii) of theorem \ref{thm6} is to 
control the number of photons in the ground state $\psp$ uniformly with 
respect to $\sigma$. The operator number of photons
$N_{ph}$ is given by
$$
N_{ph}:=\sum_{\mu =1,2}\int_{\R^3}d^3 k\  a^*_{\mu}(k)a_{\mu}(k)
$$
and we set 
$$
G(k):=  \vak^{1/2}|\rho({k})|+ 
\frac{|\rho(k)|}{\vak^{1/2}}\ .
$$

\begin{lemma}\label{lem11}
There exists a constant $C$ independent of $g$ and $\sigma$ such that 
\be
\norm{(I\otimes N_{ph}^{1/2})\psp }\leq C\vag \left(\int 
\frac{|G(k)|^2}{\vak^2}d^3k \right)^{1/2}
\ee
for every $\sigma \in (0,\sigma_0 ]$, $\vag \leq g_{2}$ and $|P_{3}|\leq 
\alpha$ ($g_{2}$ and $\alpha$ are introduced in lemma \ref{lem8}).
\end{lemma}
\proof
One easily verifies that one has the following "pull through" formula
\be \label{pull}
a_{\mu}(k)\hsp=H_{\sigma}(P_{3}-k_{3})a_{\mu}(k)+\omega(k)a_{\mu}(k)+
v_{\mu}(k)
\ee
with
\begin{align}\begin{split}\nonumber
    v_{\mu}(k)&= \frac{ig}{2\pi 
    m}\vak^{1/2}\rho_{\sigma}(k)e^{-ik'\cdot x'}  
    \left( \frac{k}{\vak}\wedge \epsilon_{\mu}(k)\right)\\
    &-\sum_{j=1,2}\frac{g}{2\pi 
    m}\frac{\rho_{\sigma}(k)}{\vak^{1/2}}e^{-ik'\cdot x'} 
    \epsilon_{\mu}(k)_{j}(p_{j}-ea_{j}(x'))\\
    &-\frac{g}{2\pi 
    m}\frac{\rho_{\sigma}(k)}{\vak^{1/2}}e^{-ik'\cdot x'} 
    \epsilon_{\mu}(k)_{3}(\pp-d\Gamma(k_{3}))\\
    &+\frac{g^2}{2\pi 
    m}\frac{\rho_{\sigma}(k)}{\vak^{1/2}}e^{-ik'\cdot x'} 
    \epsilon_{\mu}(k)\cdot A(x',0,\rho_{\sigma}) \ .
\end{split}\end{align}
Applying \eqref{pull} to $\psp$, we obtain
\begin{align}\begin{split}\nonumber
  0=&(H_{\sigma}(P_{3}-k_{3})-\esp +\omega(k))a_{\mu}(k)\psp+
v_{\mu}(k)\psp\\
=&(H_{\sigma}(P_{3}-k_{3})-E_{\sigma}(P_{3}-k_{3})+ E_{\sigma}(P_{3}-k_{3})
-\esp +\omega(k))a_{\mu}(k)\psp\\
&+v_{\mu}(k)\psp
 \end{split}\end{align}

and thus, as  $H_{\sigma}(P_{3}-k_{3})-E_{\sigma}(P_{3}-k_{3})\geq 0$,
we get using \eqref{estimee},
\begin{align}\begin{split}\label{48}
    \norm{a_{\mu}(k)\psp}&\leq \frac{1}{|E_{\sigma}(P_{3}-k_{3})
-\esp +\omega(k)|}\norm{v_{\mu}(k)\psp}\\
&\leq \frac{4}{\vak}\norm{v_{\mu}(k)\psp}
\end{split}\end{align}
for $|\pp|\leq \alpha$ and $\vag\leq g_{2}$. Using estimates from section 
\ref{reduc} (and similarly as \eqref{estimHI}) we show that there 
exists a constant $C>0$ such that
\be \label{vmuk}
\norm{v_{\mu}(k)\psp}\leq C\vag {G(k)}(\norm{(\hop -e(b,V))\psp} +1)\ .
\ee
Now, similarly as \eqref{estimH0}, we have for $\vag\leq g_{2}$
$$
\norm{\hop \psp}\leq 2|\esp| +2Cg \ .
$$
By lemma \ref{lem7} $|\esp| \leq C\vag +\frac{\pp^2}{2m}$ and therefore
we deduce from \eqref{48} that
$$
\norm{a_{\mu}(k)\psp}\leq C\vag \frac{G(k)}{\vak}
$$
where the constant $C$ is uniform with respect to
$|\pp|\leq \alpha$, $\sigma \in (0,\sigma_0 ]$ and $\vag\leq g_{2}$. 

Thus lemma \ref{lem11} follows from this last inequality and from
$$
\norm{(I\otimes N_{ph}^{1/2})\psp}^2=\sum_{\mu =1,2}\int d^3 k
\norm{(I\otimes a_{\mu}(k))\psp}^2\ .
$$
Let us remark that the above proof is a little bit formal since we do 
not check that $\psp$ belongs to the domain of the different 
operators involved in the pull through formula \eqref{pull}. But by 
mimicking \cite{H04} one easily gets a rigourous proof. We omit the 
details.
\qed

\medskip

Recall that we denote by $P_{(.]}$ 
the spectral measure of $h(b,V)$ and by
$P_{\Omega_{ph}}$ the orthogonal projection on $\Omega_{ph}$. 
We have the following
\begin{lemma}\label{lem12}
Fix $\lambda \in (e(b,V),0)$. There exists $\delta_{g}(\lambda)>0$ 
such that $\delta_{g}(\lambda)\to 0$ when $g\to 0$ and
\be
\label{52}
\langle P_{[\lambda ,\infty)}\otimes P_{\om} \psp \ ,\ \psp
\rangle \leq \delta_{g}(\lambda)
\ee
for every $\sigma \in (0,\sigma_0 ]$, $|\pp|\leq \alpha$ and $\vag \leq g_{2}$.
\end{lemma}
\proof
Since $P_{\om}H_{ph} =0$ and $P_{\om}(P_{3}-d\Gamma(k_{3}))^2= \pp^2 
P_{\om}$ we get
\begin{align}\begin{split}\nonumber
    (&P_{[\lambda ,\infty)}\otimes P_{\om})(\hsp -\esp)= P_{[\lambda 
    ,\infty)}(h(b,V)\otimes I)\otimes P_{\om} \\
    &+ (\frac{\pp^2}{2m}-\esp)P_{[\lambda ,\infty)}\otimes P_{\om}+
    P_{[\lambda ,\infty)}\otimes P_{\om}\hisp\ .
\end{split}\end{align}
Applying this last equality to $\psp$ we obtain
\begin{align}\begin{split}\label{54}
    0&=P_{[\lambda 
    ,\infty)}(h(b,V)\otimes I)\otimes P_{\om}\psp \\
    &+ (\frac{\pp^2}{2m}-\esp)P_{[\lambda ,\infty)}\otimes P_{\om}\psp\\
    &+
    P_{[\lambda ,\infty)}\otimes P_{\om}\hisp\psp\ .
\end{split}\end{align}
Since $h(b,V)P_{[\lambda ,\infty)}\geq \lambda P_{[\lambda ,\infty)}$
we obtain from \eqref{54} and lemma \ref{lem7}
\begin{align}\begin{split}\nonumber
    &\langle P_{[\lambda ,\infty)}\otimes P_{\om} \psp \ ,\ \psp
    \rangle \leq \\
    &\frac{-1}{\lambda-e(b,V)}\langle (P_{[\lambda ,\infty)}
    \otimes P_{\om})\hisp \psp \ ,\ \psp
\rangle 
\end{split}\end{align}
for every $\sigma \in (0,\sigma_0 ]$.
The lemma then follows from \eqref{estimHI} (which is also valid for 
$\hisp$).
\qed

\medskip

We are now able to conclude the proof of (iii) of theorem \ref{thm6}.
We have
\begin{align}\begin{split}\nonumber
    \langle P_{(-\infty,\lambda ]}\otimes P_{\om} \psp \ ,\ \psp
    \rangle = 1
   & -\langle P_{[\lambda ,\infty)}\otimes P_{\om} \psp \ ,\ \psp
    \rangle \\
    &-\langle 1\otimes P^\perp_{\om} \psp \ ,\ \psp
	\rangle \ .
\end{split}\end{align}
Now it suffices to remark that the second term in the right hand side of this equality is 
estimated by lemma \ref{lem12} and, noticing that 
$||P^\perp_{\om}\phi||\leq 
||N_{ph}^{1/2}\phi ||$, the third term is estimated by lemma \ref{lem11}.
\qed

\appendix

\section[The Hamiltonian with 
infrared cutoff]{Exitences of a ground state for the Hamiltonian with 
infrared cutoff}\label{Ap}

In this appendix we prove the assertion (ii) of theorem \ref{thm6} :
for $\sigma$ and $P_{3}$ small enough, the Hamiltonian with infrared 
cutoff has a ground state. This result is not surprising but the 
complete proof is long. Actually it follows by mimicking 
\cite{FGSray,FGScom,DG99} (see also \cite{Mo})
 and, here, we only give 
a sketch of the proof.

In this appendix we are faced with the lack of smoothness of the 
$\epsilon_{\mu}(k)$'s which define  vector fields on spheres 
$|k|=$cst (see \cite{LL,G04}). It suffices to consider one example. 
From now on suppose that
$$
\epsilon_{1}(k)= \frac{1}{\sqrt{k_{1}^2+k_{2}^2}}(k_{2},-k_{1},0)\quad
\mathrm{ and }\quad \epsilon_{2}(k)= \frac{k}{\vak}\wedge \epsilon_{1}(k)\ .
$$
The functions $\epsilon_{\mu}(k)$, $\mu =1,2$, are smooth only on
$\R^3\setminus \{(0,0,k_{3})\mid k_{3}\in \R \}$. Nevertheless, in 
our case, we can overcome this problem easily by choosing the 
regularization $\rho_{\sigma}$ of $\rho$ as a $C^\infty$ function 
whose support does not intersect the line 
$\{(0,0,k_{3})\mid k_{3}\in \R \}$. From now on we suppose that it is 
the case.

Let $\omod$ be the modified dispersion relation as defined in 
(\cite{FGScom}, section 5, hypothesis 3), i.e. :
$\omod$ is a smooth function satisfying 
\begin{itemize}
  \item[(i)] $\omod \geq \max (\vak, \frac{\sigma}{2})$ for all $k\in 
  \R^3$, $\omod =\vak $ for $\vak \geq \sigma$.
  \item[(ii)] $ |\nabla \omod| \leq 1$ for all $k\in 
  \R^3$, and $\nabla \omod \neq 0$ unless $k=0$.
  \item[(iii)] $\omega_{\mathrm{mod}}(k_{1}+k_{2})\leq 
  \omega_{\mathrm{mod}}(k_{1})+ \omega_{\mathrm{mod}}(k_{2})$ for all 
  $k_{1},k_{2} \in \R^{3}$.
    \end{itemize}
    We set 
$$
H_{ph,\mathrm{mod}}= \sum_{\mu =1,2}\int \omod 
a^\star_{\mu}(k)a_{\mu}(k)d^3k
$$
and
$$
\hmsp =h(b,V)\otimes I + I \otimes \left\{ 
\frac{1}{2m}(P_{3}-d\Gamma(k_{3}))^2+ H_{ph,\mathrm{mod}} \right \}
+H_{I,\sigma}(P_{3}) \ .
$$
Theorem \ref{thm3}, with the same assumption \eqref{rhosa}, is still 
valid for $\hmsp$. Set $\emsp := \inf \sigma (\hmsp)$. Then $\emsp$ 
still satisfies lemma \ref{lem8} and \eqref{estimee} for the same 
constants $g_{2}$ and $\alpha$. Moreover, according to (\cite{FGScom}; 
thm 3), $\esp =\emsp$ for $|\pp|\leq \alpha$ and $\vag\leq g_{2}$ 
and $\esp$ is an eigenvalue of $\hsp$ if and only if $\emsp$ is an eigenvalue 
of $\hmsp$. Thus in order to prove that $\hsp$ has a ground state it 
suffices to prove that $\emsp < \inf \sigma_{\mathrm{ess}}(\hmsp)$. 
The proof is by contradiction, so we suppose that 
$\emsp = \inf \sigma_{\mathrm{ess}}(\hmsp)$ and we set
\be \label{B3}
\lambda =\emsp = \inf \sigma_{\mathrm{ess}}(\hmsp) \ .
\ee
We now observe that $\emsp$ satisfies \eqref{estime} for $\vag \leq g_{2}$. 
\\
Let $\delta := \mathrm{dist}\left(e(v,V)\ ,\ \sigma(h(b,V))\setminus \{ 
e(b,V)\}\right) >0$. According to  \eqref{estime} there exist $0<\beta\leq 
\alpha$ and $0<g_{3}\leq g_{2}$ such that 
\begin{align}\begin{split}
    &\lambda \leq e(b,V)+\frac{\delta}{3}\quad \mathrm{ for } 
    |\pp|\leq \beta\\
    &C\vag \leq \frac{\delta}{12}\quad \mathrm{ for } \vag\leq g_{3}
\end{split}\end{align}
where $C$ is the constant in \eqref{estime}. 

Let $\Delta$ be an interval such that $\lambda\in \Delta$ and $\sup 
\Delta < e(b,V)+\frac{\delta}{2}$. Thus
$$
e(b,V)+\frac{2\delta}{3}-\sup \Delta -C\vag \geq \frac{\delta}{12}
$$
for $|\pp|\leq \beta$ and $\vag\leq g_{3}$ and we introduce 
$\eta>0$ such that
$$
\eta^2<e(b,V) +\frac{2\delta}{3}-\sup \Delta -C\vag\ .
$$
Then, along the same lines as in the proof of theorem II.1 in 
\cite{BFS98b}, one easily shows that it exists $M_{\Delta}$ such that 
for any $|\pp|\leq \beta$ and any $\vag\leq g_{3}$
\be \label{B9}
\norm{(e^{\eta |x'|}\otimes I)\chi_{\Delta}(\hmsp)}\leq M_{\Delta}\ .
\ee
Since we assume $\lambda \in \sigma_{\mathrm{ess}}(\hmsp)$ there exits a 
sequence $(\phi_{n})_{n\geq 1}$, with $\|\phi_{n}\|=1$, such that
\begin{align}\begin{split}\nonumber
   &\phi_{n} \in \mathrm{Ran }\chi_{\Delta}(\hmsp)\ ,\\
   &(\hmsp -\lambda)\phi_{n}\rightarrow_{n\to 0}0 ,\\
   &w-\lim_{n\to 0} \phi_{n} =0 
\end{split}\end{align}
and therefore
\be \label{lamb}
\lambda = \lim_{n\to \infty} \langle \phi_{n}, \hmsp \phi_{n}\rangle\ .
\ee
Notice that, as for \eqref{B9}, one easily shows that for any $|\pp|\leq \beta$,
any 
$\vag\leq g_{3}$ and any $n\geq 1$
\begin{align}\begin{split}
   &\norm{(e^{\frac{\eta}{2} |x'|}\otimes I)\chi_{\Delta}(\hmsp) \nabla_{x'} 
   \phi_{n}}\leq M'_{\Delta}\\
   &\norm{(e^{\frac{\eta}{2} |x'|}\otimes I)\chi_{\Delta}(\hmsp) d\Gamma(k_{3}) 
      \phi_{n}}\leq M'_{\Delta}
\end{split}\end{align}
where $M'_{\Delta}$ is a finite constant.
 
Now, in order to estimate $\langle \phi_{n}, \hmsp \phi_{n}\rangle$ 
from below, we need to localize the photons. Let $j_{0},j_{\infty}\in C^\infty(\R^3)$ be real valued functions 
with $j_{0}^{2}+j_{\infty}^2=1$, $j_{0}(y)=1$ for $|y|\leq 1$ and 
$j_{0}(y)=0$ for $|y|\geq 2$. Given $R>0$, we set 
$j_{\cdot,R}(y)=j_{\cdot}(\frac{y}{R})$ and let 
$j_{R}=(j_{0,R},j_{\infty,R})$. Here $y=\frac{1}{i}\nabla_{k}$ and 
$j_{R}$ is an operator from $\F$ to $\F \otimes \F$.

Then let $\check{\Gamma}(j_{R})$, $d\check{\Gamma}(j_{R}, \underline \omo 
j_{R}-j_{R}\omo)$ be the operators from $L^2(\R^3,\C^2)\otimes \F$ to 
$L^2(\R^3,\C^2)\otimes \F\otimes\F$ as defined in sections 2.13 and 
2.14 of \cite{DG99} (see also section 2.6 of \cite{FGScom}). Here 
$\underline \omo :=(\omo,\omo)$. Roughly speaking, 
$\check{\Gamma}(j_{R})$ separates the set of photons between photons 
localized around the electron and photons that escape to infinity 
(when $R\to \infty)$.

Set 
\begin{align}\begin{split}\nonumber
   &G_{l,\mu}(x',\rho_{\sigma})=\frac{1}{2\pi}\frac{\rho_{\sigma}}
   {\vak^{1/2}}e^{-ikx'} \epsilon_{\mu}(k)_{l}\ ,\quad l=1,2,3,\\
   &H_{\mu}(x',\rs)=-\frac{i}{2\pi}\vak^{1/2}\rs(k) \sigma .(\frac{
   k}{\vak}\wedge \epsilon_{\mu}(k))e^{-ikx'}\ ,\\
   &\Phi_{\mu}(h)=\frac{1}{\sqrt 2}(a_{\mu}(h) +a_{\mu}^\star (h)) \ .
\end{split}\end{align}

Let $\htmsp$ be the following operator in $L^2(\R^3,\C^2)\otimes 
\F\otimes\F$ :
\begin{align}\begin{split}\nonumber
    \check{H}&_{\mathrm{mod},\sigma}(P_{3})=h(b,V)\otimes I\otimes I +I\otimes 
   H_{\mathrm{mod},ph}\otimes I +I\otimes I\otimes H_{\mathrm{mod},ph}\\
   +&\frac{1}{2m}\ I\otimes (\pp -I\otimes d\Gamma(k_{3})\otimes I-I\otimes I\otimes 
   d\Gamma(k_{3}))^2\\
   -&\frac{g}{2m}\sqrt 2 \sum_{\mu 
   =1,2}\Phi_{\mu}(H_{\mu}(x',\rs))\otimes I \\
   -& \frac{g}{m}\sqrt 2 \sum_{\mu 
   =1,2}\sum_{j 
   =1,2}(p_{j}-ea_{j}(x'))\Phi_{\mu}(G_{j,\mu}(x',\rho_{\sigma}))\otimes 
   I\\
-&\frac{g}{m}\sqrt 2 (\pp -I\otimes d\Gamma(k_{3})\otimes I-I\otimes I\otimes 
   d\Gamma(k_{3})) \left( \sum_{\mu 
   =1,2} \Phi_{\mu}(G_{3,\mu}(x',\rho_{\sigma}))\otimes I\right)\\
 +&\frac{g^2}{2m}\sum_{l=1,2,3}\left(\sum_{\mu =1,2}\Phi_{\mu}
 (G_{l,\mu}(x',\rho_{\sigma}))\right)^2\otimes I  \ .
\end{split}\end{align}
Again, assuming \eqref{rhosa}, $\htmsp$ is a self-adjoint operator in 
$L^2(\R^3,\C^2)\otimes 
\F\otimes\F$ for $\vag\leq g_{1}$. We remark that 
\be \label{htmsp}
\htmsp \geq \emsp +I\otimes I\otimes H_{\mathrm{mod},ph}
\ee
and thus $\langle \phi, \htmsp \phi\rangle \geq \emsp 
+\frac{\sigma}{2}$ if the state $\phi$ has a component along the 
delocalized photons. Actually we are going to prove that, since 
$\lambda$ is in the essential spectrum, $\check{\Gamma}(j_{R})\phi_{n}$ has a non vanishing 
component along the delocalized photons and thus, in view of 
\eqref{lamb}, we will obtain a contradiction with \eqref{B3}.

Finally set for $l=1,2,3$
$$
T_{l,R}(G)(x')=\sum_{\mu=1,2}\left( \Phi_{\mu}((j_{0,R}-1)
G_{l,\mu}(x',\rho_{\sigma}))\otimes 1 + 1\otimes \Phi_{\mu}(j_{\infty,R}
G_{l,\mu}(x',\rho_{\sigma}))\right)
$$
and
$$
T_{R}(H)(x')=\sum_{\mu=1,2}\left( \Phi_{\mu}((j_{0,R}-1)
H_{\mu}(x',\rho_{\sigma}))\otimes 1 + 1\otimes \Phi_{\mu}(j_{\infty,R}
H_{\mu}(x',\rho_{\sigma}))\right)\ .
$$
By using \cite{DG99} (sections 2.13 and 2.14) and \cite{FGScom} 
(sections 2.5 and 2.6) we obtain that $\htmsp$ and $\hmsp$ are almost conjugated by $\check{\Gamma}(j_{R})$, namely
\begin{align}\begin{split}\label{feo}
\check{\Gamma}(j_{R})&\hmsp -\htmsp \check{\Gamma}(j_{R}) =
-\frac{g}{2m}\sqrt 2 T_{R}(H)(x')\check{\Gamma}(j_{R}) \\
-&\frac{g}{2m}\sqrt 2 (\pp -I\otimes d\Gamma(k_{3})\otimes I-I\otimes I\otimes 
  d\Gamma(k_{3}))T_{3R}(G)(x')\check{\Gamma}(j_{R}) \\
  -& \frac{g}{m}\sqrt 2 \sum_{j 
  =1,2}(p_{j}-ea_{j}(x'))T_{jR}(G)(x')\check{\Gamma}(j_{R})
-d\check{\Gamma}(j_{R},\underline{\omega_{\mathrm{mod}}}j_{R}-j_{R}
\omega_{\mathrm{mod}})\\
 -&d\check{\Gamma}(j_{R},\underline{k_{3}}j_{R}-j_{R}k_{3})\left( 
 \frac{1}{2m}(\pp -d\Gamma(k_{3}))+\sqrt 2 \sum_{\mu=1,2}\Phi_{\mu}(
 G_{3,\mu}(x',\rho_{\sigma}))\right)\\
 -&\frac{1}{2m}(\pp -I\otimes d\Gamma(k_{3})\otimes I-I\otimes I\otimes 
  d\Gamma(k_{3}))
  d\check{\Gamma}(j_{R},\underline{k_{3}}j_{R}-j_{R}k_{3})\\
  +&\frac{g^2}{2m}\left[\sum_{l=1,2,3}\sum_{\mu =1,2}\sum_{\mu' 
  =1,2} \right.\\
&\{a_{\mu}(j_{0,R}G_{l,\mu}(x',\rs))\otimes 1 + 1\otimes
a_{\mu}(j_{\infty,R}G_{l,\mu}(x',\rs)) \\
+& a^\star_{\mu}(j_{0,R}G_{l,\mu}(x',\rs))\otimes 1 + 1\otimes
a^\star_{\mu}(j_{\infty,R}G_{l,\mu}(x',\rs))\}\\
&\{a_{\mu'}(j_{0,R}G_{l,\mu'}(x',\rs))\otimes 1 + 1\otimes
a_{\mu'}(j_{\infty,R}G_{l,\mu'}(x',\rs))\\
+& a^\star_{\mu'}(j_{0,R}G_{l,\mu'}(x',\rs))\otimes 1 + 1\otimes
a^\star_{\mu'}(j_{\infty,R}G_{l,\mu'}(x',\rs))\}\\
-&\left. \sum_{l=1,2,3}\left( 
\sum_{\mu=1,2}\Phi_{\mu}(G_{l,\mu}(x',\rs))\right)^2\otimes I
  \right]\check{\Gamma}(j_{R})  \ .
\end{split}\end{align}
Since $\rs$ is a $C_{0}^\infty$ function, one has for $\gamma>0$
\begin{align}\begin{split}\nonumber
&(1-\Delta_{k'})^\gamma 
\frac{\rs(k)}{\vak^{1/2}}\epsilon_{\mu}(k)_{l} \ \in \ L^2(\R^3)\quad 
l=1,2,3,\ \mu=1,2,\ \sigma>0\\
&(1-\Delta_{k'})^\gamma 
{\rs(k)}{\vak^{1/2}}\sigma\cdot\left(\frac{k}{\vak}\wedge\epsilon_{\mu}(k)\right)
\ \in \ L^2(\R^3)\quad 
 \mu=1,2,\ \sigma>0\ .
\end{split}\end{align}
Here $k'=(k_{1},k_{2})$.

We then prove, as in (\cite{FGScom} lemma 9), that both
$$
e^{-\frac{\eta}{2} |x'|}T_{l,R}(G)(x')(I+I\otimes N_{ph}\otimes I +I\otimes 
I\otimes N_{ph})^{-1/2}
$$
and
$$
e^{-\frac{\eta}{2} |x'|}T_{R}(H)(x')(I+I\otimes N_{ph}\otimes I +I\otimes 
I\otimes N_{ph})^{-1/2}
$$
tend to zero in $L^2(\R^3,\C^2)\otimes 
\F\otimes\F$ when $R\to \infty$. Therefore it follows from 
\eqref{B9} and \eqref{feo} that
\be \label{B24}
\langle \phi_{n},\hmsp \phi_{n}\rangle =
\langle \phi_{n},\check{\Gamma}(j_{R})^\star \htmsp \check{\Gamma}(j_{R})
\phi_{n}\rangle
+o(R^0)
\ee
uniformly in $n$ (cf. \cite{FGScom} and \cite{DG99}).

Denoting by $P_{\Omega_{\infty}}$ the orthogonal projection on the 
vacuum of delocalized photons, we have using \eqref{htmsp}
\begin{align}\nonumber
\langle \phi_{n},\check{\Gamma}(j_{R})^\star \htmsp \check{\Gamma}(j_{R})
\phi_{n}\rangle &\geq \emsp +\frac{\sigma}{2}\\ \nonumber
&-\frac{\sigma}{2} 
\langle \phi_{n},\check{\Gamma}(j_{R})^\star (I\otimes I\otimes
P_{\Omega_{\infty}})\check{\Gamma}(j_{R})
\phi_{n}\rangle\ .
\end{align}
On the other hand we verify
$$
\check{\Gamma}(j_{R})^\star (I\otimes I\otimes
P_{\Omega_{\infty}})\check{\Gamma}(j_{R})={\Gamma}(j_{0,R}^2)\ .
$$
Then, by using lemma \ref{lem7} for $\emsp$ and the compactness of\\ 
$\chi_{\Delta}(\hmsp)e^{-\eta |x'|}{\Gamma}(j_{0,R}^2)(\hmsp +i)^{-1}$ 
(see \cite{DG99} lemma 34 or \cite{FGScom} lemma 36 and \cite{AHS78b} 
theorem 2.6), we deduce from 
\eqref{B24}, letting $n\to \infty$,
$$
\lambda \geq \emsp +\frac{\sigma}{2}+o(R^0) \ .
$$
Letting $R\to \infty$ we get a contradiction with \eqref{B3} and thus 
assertion (ii) of theorem \ref{thm6} is proved.
\qed

%\bibliography{qed}
%\bibliographystyle{abbrv}
%{unsrt}

\end{document}